\documentclass{aa}
\usepackage{graphicx}
\usepackage{amsmath}
\usepackage{geometry}
\usepackage[varg]{txfonts}

\begin{document}

\title{Traveling solar-wind bulk-velocity fluctuations and their effects on electron heating in the inner heliosphere} 

\titlerunning{Bulk-velocity fluctuations and electron heating in the inner heliosphere}
\authorrunning{Fahr, Chashei, \& Verscharen}

\author{Hans J. Fahr\inst{1}
  \and Igor V. Chashei\inst{2} 
\and Daniel Verscharen\inst{3}} 

\offprints{Hans J. Fahr, \email{hfahr@astro.uni-bonn.de}}

\institute{Argelander Institute for Astronomy, University of Bonn, Auf dem H\"{u}gel 71, 53121 Bonn, Germany
  \and Lebedev Physical Institute, Leninskii Prospect 53, 117924 Moscow, Russia
  \and Space Science Center, University of New Hampshire, 8 College Road, Durham, NH 03824, USA}

\abstract{In a recent publication, we have shown that ambient plasma electrons undergo strong heating in regions associated with compressive traveling interplanetary solar-wind
bulk-velocity jumps $\Delta U$ due to their specific interactions with the jump-inherent
electric fields. After thermalization of this energy gain per shock passage
through the operation of the Buneman instability, strong electron heating occurs that substantially influences the radial electron temperature profile. While our previous study describes 
the resulting electron temperature assuming that the amplitude of the
traveling velocity jump remains constant with increasing solar distance, we now aim at a more consistent view, describing the change of the
jump amplitude with distance due to the heated electrons. We describe the
reduction of the jump amplitude due to energy expended by
the traveling jump structure. We consider three effects; namely
energy loss due to heating of electrons, energy loss due to work done
against the pick-up-ion pressure gradient, and an energy gain due to nonlinear
jump steepening. Taking these effects into account, we show 
that the decrease in jump amplitude with solar distance is more pronounced when
 the initial jump amplitude is higher in the inner solar system.
Independent of the initial jump amplitude, it eventually decreases with increasing
distance to a value of the order of $\Delta U/U\simeq 0.1$ at the position of the heliospheric termination shock, where $\Delta U$ is the jump amplitude, and $U$ is the average solar-wind bulk velocity.The electron temperature, on the other hand, is strongly
correlated with the initial jump amplitude, leading to 
electron temperatures between 6000\,K and 20\,000\,K at distances beyond 50
AU. We compare our results with in-situ measurements of the electron-core temperature from the Ulysses spacecraft in the plane of the ecliptic for $1.5\,\mathrm{AU}\leq r\leq 5\,\mathrm{AU}$, where $r$ is the distance from the Sun. We find a very good agreement between our results and these observations, which corroborates our extrapolated predictions beyond $r=5\,\mathrm{AU}$.}


\maketitle

\section{Introduction}

The electron temperature in the solar wind is expected to rapidly drop off
with increasing distance $r$ from the Sun, as soon as the electron
heat conduction serving as the prime energy source has died out
\citep{feldman75,pilipp87,scime94}. At
distances smaller than 5 AU, electron distribution functions have been
identified as core-halo structured distributions with an electron heat flux
falling off with a power law according to $\propto r^{-2.36}$ \citep[see][]{mccomas92}. On the basis
of electron data taken from the Helios, Wind, and Ulysses spacecraft, \citet{maksimovic05}
have carefully analyzed the radial change of the core-halo-strahl structure
of the electron distribution function with distance from the Sun in the range between
0.3 AU and 1.5 AU. These authors find that, while the relative abundance of
core electrons remains fairly constant with distance, the relative abundance
of halo electrons increases and that of strahl electrons decreases, suggesting that the relative increase in halo electrons is connected to the
relative loss in strahl electrons. Interestingly enough, however, both the
core electron temperature and the halo electron temperature decrease with
distance. This effect can be represented by kappa distribution
functions with decreasing kappa-indices and will be best fitted by the fall-off of the electron kappa index from $\kappa =6$ at $r=0.5\,\mathrm{AU}$ to $\kappa =3$ at $r=1.5\,\mathrm{AU}$. The increase in
the relative abundance of the halo population is interpreted as the consequence of an
isotropization of the strahl population, leading to a conversion into the
halo population \citep[see also][]{stverak09}. 

Beyond the outer ranges of the Ulysses trajectory (i.e., at
solar distances beyond 5 AU),  measurements of low-energy solar-wind electrons are not available. Up to now, electron temperatures have been expected to fall off to negligible values in this region for theoretical reasons. At such
large distances from the Sun, processes like whistler-wave-turbulence generation due to
instabilities driven by the electron heat flux \citep[see][]{scime94,gary94} become unimportant. Also pitch-angle scattering and energy-diffusion processes can be neglected at those distances \citep{schlickeiser91,achatz93}. However, more recently
\citet{breech09} have presented a theoretical study of the heating of
solar-wind protons and electrons by dissipation of MHD-turbulent energy.
While their study shows that the theoretically obtained proton temperatures fit the Ulysses data, the theoretical electron temperatures \citep[see Fig.~3 in][]{breech09}, since being too low, miss almost all the data.

\begin{figure}
\resizebox{\hsize}{!}{\includegraphics{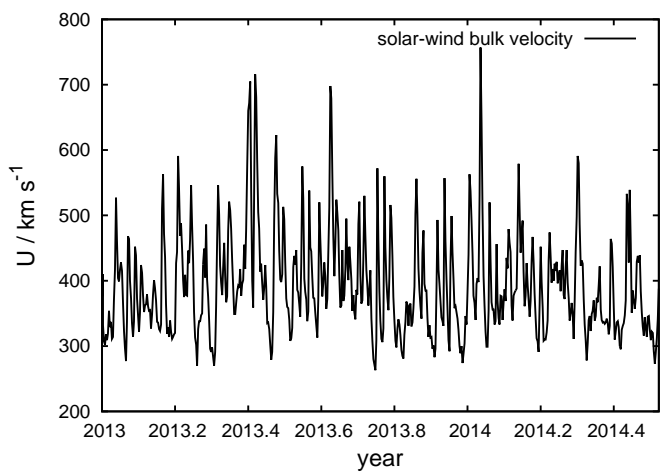}}
\caption{Solar-wind bulk velocity as a function of time in the plane of the ecliptic at 1 AU. We show OMNI-2 data (from ftp://spdf.gsfc.nasa.gov/pub/data/omni/low\_res\_omni/)  from a combined set of spacecraft observations to demonstrate the ubiquitous fluctuations in the solar-wind bulk velocity.}
  \label{figure1}
\end{figure}

As a remedy of that failure, we most recently conjectured that
the interaction between electrons and the electric fields associated with traveling fluctuations in the solar-wind bulk velocity (i.e., traveling shocks) can provide an energy source for
electron heating in this part of the heliosphere \citep{chashei14}. All solar-wind properties, including the solar-wind bulk velocity $U$, show strongly pronounced variations on many time scales as well as shock-like structures  \citep{feng09,yue11,janvier14}. We show a time line of the measured solar-wind bulk velocity in the plane of the ecliptic at 1 AU in Fig.~\ref{figure1} \citep[cf][]{echer05,lai10,sokol13}. In agreement with these observations, we find a typical occurrence rate of about 30 jumps of significant amplitude per year. These jumps are convected over the spacecraft with an average solar-wind bulk velocity of $U\approx 400\,\mathrm{km/s}$, leading to a typical distance of about $L_{\mathrm j}=3\,\mathrm{AU}$ between subsequent shocks.

Fluctuations  $\Delta U(t)\equiv U(t)-\left\langle U(t)\right\rangle $, where $\left\langle\ \cdot \right \rangle$ is the time average and $U$ is the solar-wind bulk velocity, persist to large 
distances from the Sun up to 20 to 40 AU as clearly demonstrated by Voyager observations \citep{richardson95}. Their Fig.~2 shows
that, while the bulk velocity fluctuations survive up to large solar
distances, the average bulk velocity $U=\left\langle
U(t)\right\rangle$ appears to be constant, implying that differential
kinetic energy is not converted into kinetic energy of the wind,
but into thermal degrees of freedom of the plasma system.

As mentioned in the beginning, we suspect that these  bulk-velocity fluctuations
are responsible for the still not well understood heating of electrons at larger distances from the Sun. We have recently proposed in a quantitative discussion that compressional bulk-velocity waves heat solar-wind electrons \citep{chashei14}. In this latter work, we determined the
fraction of the differential kinetic energy of the traveling shocks and quantified the energy that is transferred to thermal energy of the solar-wind electrons by means of the Buneman instability as a function of the bulk speed $U_{2}$ downstream of the velocity jump.  The joint bulk speed of electrons and protons, after passing the jump-associated electric-field jump, is given by
\begin{equation}
m_{\mathrm e}U_{2\mathrm e}+m_{\mathrm p}U_{2\mathrm p}=\left(m_{\mathrm e}+m_{\mathrm p}\right)U_{2},
\end{equation}
where $m_{\mathrm e,\mathrm p}$ denote the mass of the electron and of the proton,
respectively, and $U_{2}$ denotes the bulk velocity of the downstream
center-of-mass system. With $m_{\mathrm e}\ll m_{\mathrm p}$, this expression leads to \citep[see][]{chashei14} 
\begin{equation}
U_{2}\approx \ U_{2\mathrm p}+\frac{m_{\mathrm e}}{m_{\mathrm p}}U_{2\mathrm e}\approx \ U_{2\mathrm p}+s\sqrt{\frac{m_{\mathrm e}}{m_{\mathrm p}}}U_{2\mathrm p},
\end{equation}
where $s\equiv(U+\Delta U)/(U-\Delta U)$ is the jump compression ratio. The difference $U_{2}-U_{2\mathrm p}$ is very small compared to $U_{2\mathrm p}$, and hence the overshoot energy of the electrons in the downstream bulk frame is given by
\begin{equation}
\Delta W_{\mathrm e}=\frac{1}{2}\frac{m_{\mathrm e}m_{\mathrm p}}{m_{\mathrm e}+m_{\mathrm p}}\left(U_{2\mathrm e}-U_{2\mathrm p}\right)^{2}\approx \frac{1}{2}m_{\mathrm e}U_{2\mathrm e}^{2}.
\end{equation}
If this kinetic energy $\Delta W_{\mathrm e}$ of the overshooting electrons can
 be locally converted into electron heat, this process leads to an
electron temperature increase $\Delta T_{\mathrm e}$ after each jump passage given by
\begin{equation}\label{tjump}
\Delta T_{\mathrm e}=\frac{m_{\mathrm e}U_{2\mathrm e}^{2}}{3k}=\frac{m_{\mathrm p}\Delta U^{2}}{3k}\left(1-\frac{1}{s^{2}}\right),
\end{equation}
where $k$ is the Boltzmann constant.

This process describes an average gain of thermal energy that leads to a systematic heating of the solar-wind electrons per radial increment $\mathrm dr$
due to repeated shock passages. The resulting radial dependence of the
electron temperature can be described by a transport equation for the thermal energy. 
We expect that the electron heating due to accumulated jump passages in the heliosphere beyond about 5 AU is statistical in nature. We denote the
average distance between consecutive jumps as $L_{\mathrm j}$ and define the 
average jump occurrence rate as $\nu _{\mathrm j}\equiv U/L_{\mathrm j}$. With these
definitions, the equation for the radial electron temperature is given in
the following differential form \citep[see][]{chashei14}:
\begin{equation}\label{dTe}
\frac{\mathrm dT_{\mathrm e}}{\mathrm dr}+2\frac{T_{\mathrm e}}{r}=\Delta T_{\mathrm e}\frac{\Delta U}{L_{\mathrm j}U}=\Delta T_{\mathrm e}\frac{\Delta X}{L_{\mathrm j}},
\end{equation}
where $\Delta X\equiv\Delta U/U$.
The term on the right-hand side of Eq.~(\ref{dTe}) describes the electron heating
induced by jump passages. \citet{chashei14}, when solving the above
equation, assumed that $\Delta T_{\mathrm e}=\Delta T_{\mathrm e}(\Delta U)$ is a
constant. This assumption is true if $\Delta U$ is independent of the
distance $r$. In that case, the radial profile of the resulting electron
temperature is given by 
\begin{equation}
T_{\mathrm e}(x)=\frac{1}{x^{2}}\left(x^{3} \frac{m_{\mathrm p}U^{2}\Delta X^{3}}{18k}\frac{r_{0}}{L_{\mathrm j}}+T_{\mathrm e0}\right),
\end{equation}
where $x\equiv r/r_{0}$ is the dimensionless spatial coordinate and $T_{\mathrm e0}$ denotes the electron temperature at $r=r_{0}=1\,\mathrm{AU}$ \citep[solution shown in Fig.~1 of][]{chashei14}. 
This solution suffers from the inconsistency that the jump kinetic energy is assumed to be constant even after transferring energy to the electrons. In the following section of this study, we
make this earlier approach more consistent by taking into account the energy consumption at the passage of each jump during this process.

\section{Change of the Jump Amplitude with Distance from the Sun}

In order to increase the consistency of our approach, we now include higher-order corrections to the electron heating due to the variation of the jump amplitude $\Delta U$ with distance $r$. This amplitude is assumed to be the primary physical reason for the gain of thermal energy of the electrons. Therefore, we have to describe the change of $\Delta U$ due to energy 
expended by the excess kinetic energy of the jump structure adequately. In-situ observations by the Voyager-2 spacecraft at distances between 10 AU and 40 AU from the Sun \citep{richardson95} show that, compared to solar-wind bulk-velocity measurements carried out simultaneously at 1 AU
by IMP-8, the average solar-wind speed does not change with distance. On the other hand, the amplitude of the speed fluctuations strongly decreases with distance from the Sun \citep[see Fig.~6 of][]{richardson95}.
This observation indicates that these fluctuations do work, while the bulk-solar-wind outflow does not. Our theoretical approach is based on these observations, adopting that the average solar-wind speed $U $ is constant with distance from the Sun.
Based on this observationally supported assumption, we consider three effects that determine the change of $\Delta U$ with distance $r$:
\begin{enumerate}[a)]
\item heating of electrons,\\
\item work done against the slower side of the jump with its higher pick-up-ion
pressure, and\\
\item steepening of the jump profile by nonlinear superpositions of small-scale
bulk-velocity fluctuations.
\end{enumerate}
In the following we shall separately look into these three different
effects.

\subsection{a) Reduced Compression due to Electron Heating}\label{sect:a}

We consider the spatial divergence of the jump-associated flow of excess kinetic
 energy on the high-velocity side of a jump with the amplitude $\Delta X=\Delta U/U$.
This jump acts as a local source of electron thermal energy, and this
heating reflects a local energy sink for the excess kinetic
energy that is represented by the compression profile $\Delta X(r)$. Using Eq.~(\ref{dTe}) for the electron temperature, we can formulate an expression for the energy sink associated with this jump as the
divergence of the excess kinetic energy flow:
\begin{equation}
\frac{1}{r^{2}}\frac{\mathrm d}{\mathrm dr}r^{2}\left(U\frac{1}{2}n_{\mathrm e}m_{\mathrm p}\,\Delta U^{2}\right)=-\frac{3}{2}Uk\,\Delta T_{\mathrm e}\,n_{\mathrm e}\frac{\Delta U}{L_{\mathrm j}U},
\end{equation}
where $n_{\mathrm e}=n_{\mathrm p}=n$ is the local solar wind electron/proton number density and $\Delta T_{\mathrm e}$ is the electron temperature increase per jump passage as
given by Eq.~(\ref{tjump}). We assume that the mean bulk
velocity $U=(1/2)(U+\Delta U+U-\Delta U)$ is constant and find
\begin{equation}\label{divergence}
\frac{\mathrm d}{\mathrm dr}\left(\Delta U^{2}n_{\mathrm e}\right)+\frac{2}{r}\left(\Delta U^{2}n_{\mathrm e}\right)=-3\frac{k\,\Delta T_{\mathrm e}}{m_{\mathrm p}U}\frac{\Delta U}{L_{\mathrm j}}.
\end{equation}
Supported by Voyager data, we assume that the traveling jumps in bulk
velocity have a small amplitude ($\Delta U\ll U$, which is equivalent to $\Delta X \ll 1$).
This observation allows us to approximate the electron-temperature increase per jump passage in Eq.~(\ref{tjump}) using the linearizations
\begin{equation}
s=\frac{1+\Delta X}{1-\Delta X}\simeq 1+2\,\Delta X
\end{equation}
and
\begin{equation}
s^{2}\simeq (1+2\,\Delta X)^{2}\simeq 1+4\,\Delta X.
\end{equation}
We can then rewrite Eq.~(\ref{tjump}) as
\begin{equation}\label{DeltaTe}
\Delta T_{\mathrm e}\simeq \frac{m_{\mathrm p}\Delta U^{2}}{3k}\left[1-(1-4\,\Delta X)\right]=\frac{4m_{\mathrm p}U^{2}\,\Delta X^{3}}{3k}
\end{equation}
and obtain from Eq.~(\ref{divergence})
\begin{equation}
\frac{2}{\Delta X}\frac{\mathrm d\,\Delta X}{\mathrm dr}+\frac{1}{n_{\mathrm e}}\frac{\mathrm dn_{\mathrm e}}{\mathrm dr}+\frac{2}{r}=-4\frac{\Delta X^{2}}{L_{\mathrm j}}.
\end{equation}
Assuming a spherically symmetric decrease in density of the average solar wind flow with $n_{\mathrm e}\propto r^{-2}$, we then obtain
\begin{equation}
\frac{2}{\Delta X}\frac{\mathrm d\,\Delta X}{\mathrm dr}=-4\frac{\Delta X^{2}}{L_{\mathrm j}}
\end{equation}
and find
\begin{equation}
\frac{\mathrm d\,\Delta X^{-2}}{\mathrm dr}=4\frac{1}{L_{\mathrm j}}.
\end{equation}
From this relation, we derive in a first step 
\begin{equation}
\left| \Delta X^{-2}\right| _{r_{0}}^{r}=4\frac{1}{L_{\mathrm j}}\int\limits_{r_{0}}^{r}\mathrm dr=\frac{4}{L_{\mathrm j}}(r-r_{0}),
\end{equation}
which delivers a solution of the form
\begin{equation}\label{deltax}
\Delta X=\frac{1}{\sqrt{\Delta X_{0}^{-2}+\frac{4}{L_{\mathrm j}}(r-r_{0})}}=\frac{\Delta X_{0}}{\sqrt{1+\frac{4r_{0}}{L_{\mathrm j}}\Delta X_{0}^{2}(x-1)}}.
\end{equation}
\begin{figure}
\resizebox{\hsize}{!}{\includegraphics{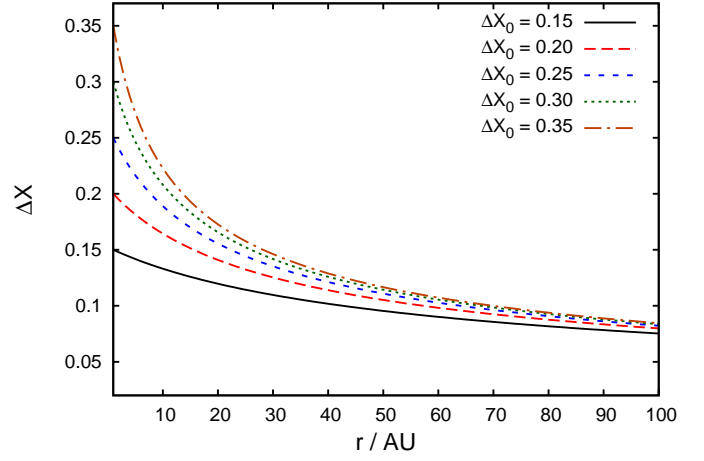}}
\caption{Compression $\Delta X$ as a function of distance $r$ from the Sun for five different values of $\Delta X_0$ at $r_0=1\,\mathrm{AU}$ with $r_0/L_{\mathrm j}=1/3$. The compression decreases with distance and approaches a value of about 0.1 at the position of the termination shock. The dependence on the jump occurrence $r_0/L_{\mathrm j}$ is discussed in \citet{chashei14}.} 
  \label{figure2}
\end{figure}
We show $\Delta X$ as a function of distance $r$ from the Sun for five different values of $\Delta X_0$ in Fig.~\ref{figure2}. We use $r_0/L_{\mathrm j}=1/3$ in agreement with observations at 1 AU (see Fig.~\ref{figure1}). The jump amplitude $\Delta X$  remarkably
decreases with increasing distance from the Sun. The decrease is even more
pronounced for cases in which the initial value $\Delta X_0$ is higher in the inner solar
system. However, independent of $\Delta X_0$, the jump amplitude assumes values of  $\lesssim 0.1$ at the position of the heliospheric termination shock (i.e., at $r\approx 90\,\mathrm{AU}$).

\begin{figure}
\resizebox{\hsize}{!}{\includegraphics{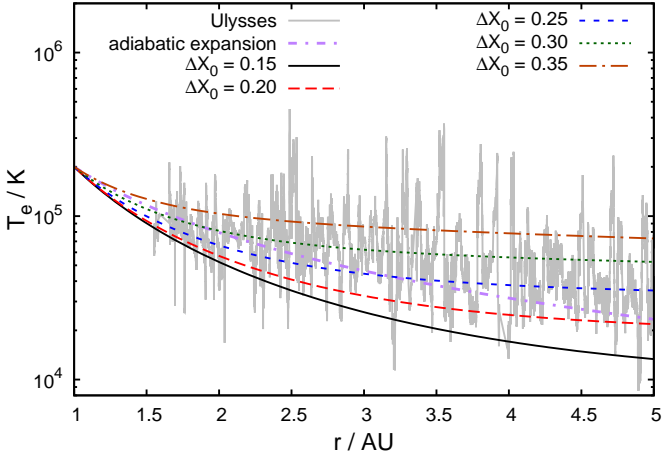}}
\caption{Electron temperature $T_{\mathrm e}$ as a function of distance $r$ from the Sun for five different values of $\Delta X_0$ with $r_0/L_{\mathrm j}=1/3$ and $U=400\,\mathrm{km/s}$. The electron temperature is greater than adiabatic. In addition, we show Ulysses observations of the in-situ electron-core temperature in the plane of the ecliptic and the prediction from adiabatic expansion.}
  \label{figure3}
\end{figure}

\begin{figure}
\resizebox{\hsize}{!}{\includegraphics{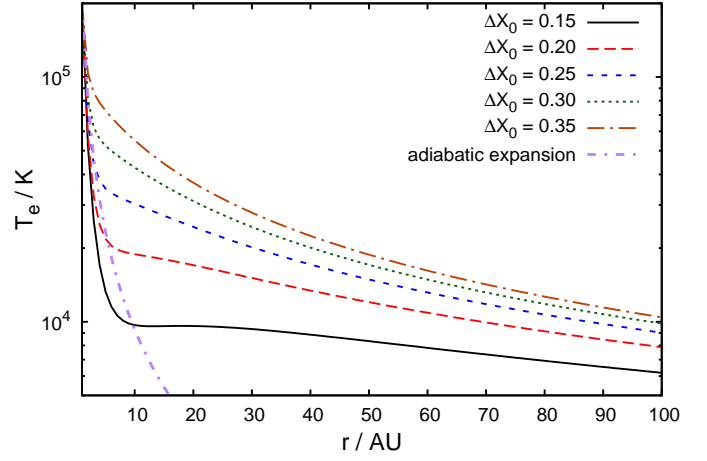}}
\caption{Electron temperature $T_{\mathrm e}$ as a function of distance $r$ from the Sun for five different values of $\Delta X_0$ with $r_0/L_{\mathrm j}=1/3$ and $U=400\,\mathrm{km/s}$. The electron temperature beyond 10 AU is higher than predicted according to adiabatic expansion.}
  \label{figure4}
\end{figure}

With this result for the dependence of $\Delta X$ on $r$, we solve the
earlier differential equation (\ref{dTe}) for the resulting electron temperature
and obtain 
\begin{equation}
\frac{\mathrm dT_{\mathrm e}}{\mathrm dx}+2\frac{T_{\mathrm e}}{x}=\frac{4}{3}\frac{m_{\mathrm p}U^{2}}{k\lambda}\frac{\Delta X_{0}^{4}}{ \left( 1+4\,\Delta X_{0}^{2}\frac{x-1}{\lambda }\right)^2},
\end{equation}
where $\lambda \equiv L_{\mathrm j}/r_{0}$.
The solution of this inhomogeneous differential equation is given by
\begin{multline}\label{eq:te}
T_{\mathrm e}(x)=\frac{1}{x^2}\left\{\frac{4}{3}\frac{m_{\mathrm p}U^{2}}{k\lambda}\Delta X_{0}^{4}\int\limits_1^x\frac{y^2}{\left[1+a(x-1)\right]^2}\mathrm dy+T_{\mathrm e0}\right\}\\
=\frac{1}{x^2}\left\{\frac{4}{3}\frac{m_{\mathrm p}U^{2}}{k\lambda a^3}\Delta X_{0}^{4}
\left[2(a-1)\ln\left[1+a(x-1)\right]\vphantom{\frac{A^2}{A^2}}\right.\right.\\
\left.\left.+\frac{a(x-1)\left[1+a(x-1)+(a-1)^2\right]}{1+a(x-1)}\right]
+T_{\mathrm e0}\right\}
\end{multline}
with $a=4\Delta X_0^2/\lambda$. We show the result of Eq.~(\ref{eq:te}) for five different values of $\Delta X_0$ in Figs.~\ref{figure3} and \ref{figure4}. We use $T_{\mathrm e0}=2\times 10^5\,\mathrm{K}$ and $U=400\,\mathrm{km/s}$. In Fig.~\ref{figure3}, we show our results for $1.5,\mathrm{AU}\leq r\leq 5\,\mathrm{AU}$ and compare them with in-situ Ulysses measurements of the electron-core temperature during the spacecraft's first orbit in the plane of the ecliptic (from December 28, 1990 until December 31, 1991). The data was taken with Ulysses' SWOOPS experiment \citep{bame92}. The comparison between our predictions and in-situ measurements shows a very good agreement between theory and observation. The modeled and observed electron temperatures are significantly higher than predicted for an adiabatically-expanding gas (i.e., $T_{\mathrm e}\propto r^{-4/3}$).
We achieve the best agreement for values of $\Delta X_0$ between 0.25 and 0.3. In Fig.~\ref{figure4}, we extrapolate our results beyond 5 AU and show our predictions for $1\,\mathrm{AU}\leq r\leq 100\,\mathrm{AU}$.
The electron temperature in our model is significantly higher than predicted according to adiabatic expansion beyond 10 AU for all shown values of $\Delta X_0$.  It assumes values of about $T_{\mathrm e}\approx 10^4\,\mathrm{K}$ at the position of the heliospheric termination shock.

\subsection{b) Change of Compression due to Work done against Entropized Pick-Up
Ions}

In this section, we consider another effect that may also contribute to a decrease in the jump
amplitude $\Delta X$, namely the work done by the faster front against the difference in 
ion pressure over the traveling shock front. The faster regime ($U_{1}=U+\Delta U$) is running into the slower regime ($U_{2}=U-\Delta U$) with a differential velocity $\Delta U$. During this process, the plasma has to do work against the pressure difference between the two regimes in order to adapt
the flow to the slower regime (i.e., $U_{2}$). We estimate the work done per unit time as
\begin{equation}
\frac{\mathrm d\epsilon (s)}{\mathrm dt}=-\frac{\Delta U\cdot \Delta P}{D}=-\Delta U\frac{P_{2}^{\ast
}-P_{1}^{\ast }}{D},
\end{equation}
where $D$ is the transit distance over the shock profile, and $P_{1,2}^{\ast
}$ are the adaptive pressures (i.e., the entropized kinetic energy densities) on
the upstream and on the downstream sides of the jump, respectively. Especially in the outer heliosphere ($r\geq 5\,\mathrm{AU}$), the ion pressure is dominated by the pick-up-ion pressures on either side of the jump. Under this assumption, the expressions for the ion pressures are significantly simplified \citep[see][]{fahr12b} for a perpendicular jump ($\Delta \vec{U}\perp \vec{B}$), leading to 
\begin{equation}
P_{2}^{\ast }-P_{1}^{\ast }\simeq P_{1,\mathrm{\mathrm{pui}}}\left[ \frac{s}{3}\left(2A_{\perp }(s)+\frac{s^{2}}{A_{\perp }^{2}(s)}\right)-1\right],
\end{equation}
where the remaining pressure adaptation function $A_{\perp }(s)$ in case
of a perpendicular shock is simply given by $A_{\perp }(s)=s$  with $s\simeq 1+2\,\Delta X$ . This leads to
\begin{multline}
P_{2}^{\ast }-P_{1}^{\ast }\simeq P_{1,\mathrm{\mathrm{pui}}}\left[\frac{s}{3}(2s+1)-1\right]\\
=P_{1,\mathrm{\mathrm{pui}}}\left[\frac{1+2\,\Delta X}{3}(3+4\,\Delta X)-1\right]\approx \frac{10}{3}\Delta X\, P_{1,\mathrm{\mathrm{pui}}},
\end{multline}
which allows us to formulate the ion-induced energy change as
\begin{multline}
\frac{\mathrm d\epsilon (s)}{\mathrm dt}=-\frac{\Delta U\cdot \Delta P}{D}=-\frac{10}{3}\frac{\Delta U}{%
D}\Delta X\, P_{1,\mathrm{\mathrm{pui}}}\\
=-\frac{10}{3}\frac{U\,\Delta X^{2}}{D}P_{1,\mathrm{\mathrm{pui}}}.
\end{multline}
With this additional term, we now obtain the following corrected differential equation for $\Delta X$:
\begin{equation}
\frac{2}{\Delta X}\frac{\mathrm d\,\Delta X}{\mathrm dr}+\frac{1}{n_{\mathrm e}}\frac{\mathrm dn_{\mathrm e}}{\mathrm dr}+\frac{2}{r}=-4\frac{\Delta X^{2}}{L_{\mathrm j}}-\frac{20}{3}\frac{P_{1,\mathrm{\mathrm{pui}}}}{n_{\mathrm e}m_{\mathrm p}U^{2}D}.
\end{equation}
We write the upstream pick-up-ion pressure in the form $P_{1,\mathrm{\mathrm{pui}}}=\zeta
n_{\mathrm e}kT_{\mathrm{\mathrm{pui}}}$ with the pick-up-ion abundance ratio $\zeta =n_{\mathrm{\mathrm{pui}}}/n_{\mathrm e}$. This leads to the new differential equation
\begin{equation}
\frac{2}{\Delta X}\frac{\mathrm d\,\Delta X}{\mathrm dr}+\frac{1}{n_{\mathrm e}}\frac{\mathrm dn_{\mathrm e}}{\mathrm dr}+\frac{2}{r}=-4\frac{\Delta X^{2}}{L_{\mathrm j}}-\frac{20}{3}\frac{1}{m_{\mathrm p}U^{2}D}\zeta kT_{\mathrm{\mathrm{pui}}}.
\end{equation}
Again, taking $n_{\mathrm e}\propto r^{-2}$, this relation then simplifies to
\begin{equation}
\frac{2}{\Delta X}\frac{\mathrm d\,\Delta X}{\mathrm dr}=-4\frac{\Delta X^{2}}{L_{\mathrm j}}-\frac{20}{3}\frac{1}{D}\frac{\zeta kT_{\mathrm{pui}}}{m_{\mathrm p}U^{2}}.
\end{equation}
We introduce the pick-up-ion Mach number $M_{\mathrm{pui}}^{2}\equiv m_{\mathrm p}U^{2}/\zeta
kT_{\mathrm{pui}}\simeq 1$ and assume that this number be constant in the outer
heliosphere \citep[see e.g.~][]{fahr99,fahr07}. We then find
\begin{multline}\label{deltaxpui}
\Delta X=\frac{1}{\sqrt{\left(\frac{3DM_{\mathrm{pui}}^2}{5L_{\mathrm j}}+\frac{1}{\Delta X_0}\right)\exp\left[\frac{20}{3}\frac{\left(r-r_0\right)}{DM_{\mathrm{pui}}^2}\right]-\frac{3DM_{\mathrm{pui}}^2}{5L_{\mathrm j}}}}\\
=\frac{\Delta X_0}{\sqrt{\left(\frac{3DM_{\mathrm{pui}}^2}{5L_{\mathrm j}}\Delta X_0^2+1\right)\exp\left[\frac{20}{3}\frac{\left(x-1\right)}{M_{\mathrm{pui}}^2}\frac{r_0}{D}\right]-\frac{3DM_{\mathrm{pui}}^2}{5L_{\mathrm j}}\Delta X_0^2}}.
\end{multline}
Expansion of the exponential term in Eq.~(\ref{deltaxpui}) for $D\ll r_0$ leads to Eq.~(\ref{deltax}). Since the shock transit distance is much smaller than 1 AU, the corrections due to the pick-up-ion pressure lead to qualitatively very similar curves as already shown in Fig.~\ref{figure2}.

\subsection{ c) Increased Compression due to Nonlinear Wave Steepening}

One may also have to envisage processes that counteract the processes described in a) and b), namely processes that support a pile-up of the bulk-velocity jump amplitude. For instance, fluctuations in the bulk velocity may cause such a pile-up by nonlinear superposition.  Therefore, we consider wave steepening in the system in addition to the previously discussed processes. Small-scale velocity fluctuations
 described by $\delta U(x,t)=\delta U(k)\cos \left[k(x-Ut)\right]$ can pile up
into a large-scale fluctuation with $L\simeq L_{\mathrm j}=2\pi /k_{\min }$ due to
nonlinear wave-coupling and dissipation processes. For one-dimensional waves, this situation is described by the following equation (see \citet{infeld90}, pp.~6-10 or \citet{treumann97}, pp.~244-280):
\begin{equation}
\frac{\partial }{\partial t}\delta U+\delta U\frac{\partial }{\partial x}\delta U=F,
\end{equation}
where $F$ denotes a dissipation force that counteracts the nonlinear term
on the left-hand side and compensates for catastrophic wave steepening and wave
breaking.
In case of the so-called Burger's equation \citep[see][]{treumann97}, a particular dissipative force is  introduced in place of $F$ that is proportional to the second derivative of the velocity perturbation, leading to the following differential equation:
\begin{equation}\label{burgers}
\frac{\partial }{\partial t}\delta U+\delta U\frac{\partial }{\partial x}\delta U=\alpha \frac{\partial ^{2}}{\partial x^{2}}\delta U,
\end{equation}
where $\alpha $ is a positive dissipation coefficient that acts like a
diffusion coefficient (assumed to be constant with distance $r$). The
background plasma moves with the velocity $U$, and $\delta U$ represents the superposition\ $\delta U=U+\Delta U$. If the nonlinear steepening of $\delta U$ (second
term on the left-hand side of the above Burger's equation (\ref{burgers})) increases, the dissipative term on the right-hand side can compensate for the nonlinear term and can allow
for a stationary solution in the system co-moving with the nonlinear wave
profile.
We assume that this developing nonlinear wave asymptotically moves with
the velocity $\Delta U$. This allows us to write the Burger's equation in
this particular co-moving system, where the first term of the left-hand side disappears (i.e., $\partial\,\delta U/\partial t=0$) when we transform the equation to space coordinates $y=x-\Delta U\, t$. This procedure then leads to 
\begin{equation}
(\delta U-\Delta U)\frac{\partial }{\partial y}\delta U=\alpha \frac{%
\partial ^{2}}{\partial y^{2}}\delta U.
\end{equation}
The solution of this equation is easily obtained in the form of a velocity shock ramp given by
\begin{equation}
\delta U-\Delta U=-\Delta U\tanh \left(\frac{\Delta U\, y}{2\alpha}\right),
\end{equation}
which can be rewritten in the form
\begin{equation}\label{tanhprofile}
\delta U=\Delta U \left[1-\tanh \left(\frac{\Delta U\, y}{2\alpha} \right)\right].
\end{equation}

In order to estimate the appropriate value of $\alpha $ (which has the dimension of cm$^{2}$/s), we return to the original Burger's equation and estimate the time scale for steepening (or in the opposite case: for dissolution) of the wave profile by the pure diffusion-type equation (i.e., domination of the
dissipation term) given by
\begin{equation}
\frac{\partial }{\partial t}\delta U=\alpha \frac{\partial ^{2}}{\partial
x^{2}}\delta U.
\end{equation}
We find the solution of this equation within the system $[-D;+D]$ by
\begin{equation}\label{duprof}
\delta U(x,t)=\delta U_{0}\frac{2D}{\sqrt{4\pi \alpha t}}\exp \left(-\frac{x^{2}}{4\alpha t}\right).
\end{equation}
The kinetic-energy density of the velocity fluctuations within the two flanks $[-D;+D]$ of such a velocity structure with the structure scale $D$ is given by
\begin{equation}
\epsilon _{\mathrm{nl}}=\frac{1}{2D}\int\limits_{-D}^{+D}\frac{1}{2}m_{\mathrm p}n_{\mathrm e}\delta U^{2}\mathrm dx.
\end{equation}
When free diffusion would operate, its temporal change is given by
\begin{equation}\label{depsilondt}
\dot{\epsilon}_{\mathrm{nl}}=\frac{1}{2D}\frac{\mathrm d}{\mathrm dt}\int\limits _{-D}^{D}\frac{1}{2}m_{\mathrm p}n_{\mathrm e}\,\delta
U^{2}\mathrm dx.
\end{equation}
Taking the above expression for $\dot{\epsilon}_{\mathrm{nl}}$ for nonlinear diffusion or
steepening per unit volume, we obtain
\begin{multline}
\dot{\epsilon}_{\mathrm{nl}}=\frac{1}{2D}\frac{1}{2}m_{\mathrm p}n_{\mathrm e}\,\delta U_{0}^{2}\frac{\mathrm d}{\mathrm dt}\left[\frac{4D^{2}}{4\pi \alpha t}\int\limits_{-D}^{D}\exp \left(-\frac{2x^{2}}{4\alpha t}\right)\mathrm dx\right]\\
=Dm_{\mathrm p}n_{\mathrm e}\,\delta U_{0}^{2}\frac{\mathrm d}{\mathrm dt}\left[\frac{\sqrt{2\alpha t}}{4\pi \alpha t}\int\limits _{-D/\sqrt{2\alpha t}}^{D/\sqrt{2\alpha t}}\exp \left(-y^{2}\right)\mathrm dy\right].
\end{multline}
Evaluating this integral expression furtheron leads to
\begin{multline}
\dot{\epsilon}_{\mathrm{nl}}=Dm_{\mathrm p}n_{\mathrm e}\,\delta U_{0}^{2}\frac{\mathrm d}{\mathrm dt}\left[\frac{\mathrm{erf}\left(\frac{D}{\sqrt{2\alpha t}}\right)}{\sqrt{8\pi \alpha t}}\right]\\
=Dm_{\mathrm p}n_{\mathrm e}\,\delta U_{0}^{2}\left[\frac{e^{-D^{2}/2\alpha t}\frac{\mathrm d}{\mathrm dt}\frac{D}{\sqrt{2\alpha t}}}{\pi \sqrt{8\alpha t}}-\frac{1}{2}\frac{\mathrm{erf}\left(\frac{D}{\sqrt{2\alpha t}}\right)}{\left(8\pi \alpha t\right)^{3/2}}8\pi\alpha \right],
\end{multline}
and finally to
\begin{equation}
\dot{\epsilon}_{\mathrm{nl}}=\frac{Dm_{\mathrm p}n_{\mathrm e}\,\delta U_{0}^{2}}{\sqrt{8\pi\alpha t}}\left[\frac{2\alpha De^{-D^{2}/2\alpha t}}{(8\pi \alpha t)^{3/2}}-\frac{\mathrm{erf}\left(\frac{D}{\sqrt{2\alpha t}}\right)}{2 t}\right].
\end{equation}
According to the profile in Eq.~(\ref{tanhprofile}), $D\simeq \alpha /\Delta U$. On the other
hand, the characteristic time $\tau $ of the shock passage is given by $\tau
=D/\Delta U=\alpha /\Delta U^{2}$. Evaluating now the above expression for this characteristic time $\tau $ leads to the following expression: 
\begin{equation}
\dot{\epsilon}_{\mathrm{nl}}=\frac{Dm_{\mathrm p}n_{\mathrm e}\,\delta U_{0}^{2}}{\sqrt{8\pi D^{2}}}\left[\frac{2D^2\,\Delta U\, e^{-D^{2}/2D^{2}}}{\left(8\pi D^2\right)^{3/2}}-\Delta U\frac{\mathrm{erf}\left(\frac{D}{\sqrt{2D^{2}}}\right)}{2 D}\right],
\end{equation}
or finally (with $\Delta U\approx \delta U_0$ as suggested by Eq.~(\ref{duprof})) to
\begin{equation}
\dot{\epsilon}_{\mathrm{nl}}=\frac{m_{\mathrm p}n_{\mathrm e}\,\Delta U^{3}}{\sqrt{8\pi}D}\left[\frac{2e^{-1/2}}{\left(8\pi\right)^{3/2}}-\frac{1}{2}\mathrm{erf}\left(\frac{1}{\sqrt{2}}\right)\right].
\end{equation}
We then obtain the following transport equation with the newly found term for $\dot{\epsilon}_{\mathrm{nl}}$:
\begin{equation}
\frac{1}{r^{2}}\frac{\mathrm d}{\mathrm dr}r^{2}\left(U\frac{1}{2}n_{\mathrm e}m_{\mathrm p}\,\Delta U^{2}\right)=-\frac{3}{2}Uk\,\Delta T_{\mathrm e}\,n_{\mathrm e}\frac{\Delta U}{UL_{\mathrm j}}-\dot{\epsilon}_{\mathrm{nl}}(\Delta U).
\end{equation}
The last term on the right-hand side represents the energy that is required in order to maintain the jump profile. Free diffusion would instead destroy the profile according to Eq.~(\ref{depsilondt}).
The transport equation is then given by
\begin{multline}
\frac{1}{r^{2}}\frac{\mathrm d}{\mathrm dr}r^{2}\left(U\frac{1}{2}n_{\mathrm e}m_{\mathrm p}\,\Delta U^{2}\right)=-\frac{3}{2}Uk\,\Delta T_{\mathrm e}\,n_{\mathrm e}\frac{\Delta U}{UL_{\mathrm j}}\\
-\frac{m_{\mathrm p}n_{\mathrm e}\,\Delta U^{3}}{\sqrt{8\pi}D}\left[\frac{2e^{-1/2}}{\left(8\pi\right)^{3/2}}-\frac{1}{2}\mathrm{erf}\left(\frac{1}{\sqrt{2}}\right)\right].
\end{multline}
With the definition
\begin{equation}
\Gamma \equiv \left[\frac{1}{2}\mathrm{erf}\left(\frac{1}{\sqrt{2}}\right)-\frac{2e^{-1/2}}{\left(8\pi\right)^{3/2}}\right]\simeq 0.3,
\end{equation}
we find 
\begin{equation}
\frac{\mathrm d}{\mathrm dr}\left(\Delta U^{2}n_{\mathrm e}\right)+\frac{2}{r}\left(\Delta U^{2}n_{\mathrm e}\right)=-3\frac{k\,\Delta T_{\mathrm e}\,n_{\mathrm e}}{m_{\mathrm p}}\frac{\Delta U}{UL_{\mathrm j}}+\frac{2n_{\mathrm e}\Delta U^{3}}{\sqrt{8\pi}DU}\Gamma.
\end{equation}
Insertion of $\Delta T_{\mathrm e}$ from Eq.~(\ref{DeltaTe}) then leads to the following differential equation:
\begin{equation}
\frac{\mathrm d}{\mathrm dr}\left(\Delta U^{2}n_{\mathrm e}\right)+\frac{2}{r}\left(\Delta U^{2}n_{\mathrm e}\right)=-\frac{4U^2n_{\mathrm e}\,\Delta X^{4}}{L_{\mathrm j}}+\frac{2n_{\mathrm e}\Delta U^{3}}{\sqrt{8\pi}DU}\Gamma,
\end{equation}
or equivalently to 
\begin{equation}
\frac{\mathrm d}{\mathrm dr}\Delta X^{2}+\Delta X^{2}\frac{1}{n_{\mathrm e}}\frac{\mathrm dn_{\mathrm e}}{\mathrm dr}+\frac{2}{r}\Delta X^{2}=-\frac{4\Delta X^{4}}{L_{\mathrm j}}+\frac{2\Delta X^{3}}{\sqrt{8\pi}D}\Gamma
\end{equation}
and
\begin{equation}
\frac{2}{\Delta X}\frac{\mathrm d}{\mathrm dr}\Delta X+\frac{1}{n_{\mathrm e}}\frac{\mathrm dn_{\mathrm e}}{\mathrm dr}+\frac{2}{r}=-\frac{4\Delta X^{2}}{L_{\mathrm j}}+\frac{2\Delta X}{\sqrt{8\pi}D}\Gamma.
\end{equation}
For a radially symmetric density drop-off, we rewrite the transport equation including terms that decrease (first term) and that increase (second term) the compression as
\begin{equation}
\frac{2}{\Delta X}\frac{\mathrm d}{\mathrm dr} \Delta X=-\frac{4\Delta X^{2}}{L_{\mathrm j}}+\frac{2\Delta X}{\sqrt{8\pi}D}\Gamma.
\end{equation}
The combination of decreasing and increasing factors can lead to a vanishing gradient and thus a constant compression $\Delta X$ if
\begin{equation}
\Delta X=\frac{\Gamma}{2\sqrt{8\pi}}\frac{L_{\mathrm j}}{D}\simeq \frac{0.3}{10}\frac{L_{\mathrm j}}{D}=3\times 10^{-2}\frac{L_{\mathrm j}}{D}.
\end{equation}
Therefore, the newly derived term for structure steepening will only
 compete with the first term in case the jump amplitude has dropped down
to values of $\Delta X\lesssim 10^{-2}(L_{\mathrm j}/D)$.
Looking for a solution of the full equation, we start from the solution of 
\begin{equation}
\frac{2}{\Delta X}\frac{\mathrm d}{\mathrm dr} \Delta X=-\frac{4\Delta X^{2}}{L_{\mathrm j}}.
\end{equation}
As shown in Section~\ref{sect:a}, the solution is given by
\begin{equation}
 \Delta X=\frac{\Delta X_{0}}{\sqrt{1+\frac{4}{L_{\mathrm j}}\Delta X_{0}^{2}\left(r-r_{0}\right)}}.
\end{equation}
The solution of the other part,
\begin{equation}
\frac{2}{\Delta X}\frac{\mathrm d}{\mathrm dr} \Delta X=\frac{2\Delta X}{\sqrt{8\pi}D}\Gamma,
\end{equation}
is derived from
\begin{equation}
\frac{1}{\Delta X^{2}}\frac{\mathrm d}{\mathrm dr} \Delta X=-\frac{\mathrm d}{\mathrm dr}\Delta X^{-1} =\frac{1}{\sqrt{8\pi}D} \Gamma 
\end{equation}
and yields
\begin{equation}
\Delta X^{-1}-\Delta X_{0}^{-1}=-\frac{1}{\sqrt{8\pi}D}\Gamma (r-r_0).
\end{equation}
This leads to
\begin{equation}
\Delta X=\frac{1}{\Delta X_{0}^{-1} -\frac{ \Gamma }{\sqrt{8\pi}D} (r-r_0)}=\frac{\Delta X_{0}}{1 -\frac{\Gamma\, \Delta X_{0}}{\sqrt{8\pi}D} (r-r_0)}.
\end{equation}
According to these considerations, the general solution is given by the superposition
\begin{equation}
\Delta X=\frac{\Omega \, \Delta X_{0}}{\sqrt{1+\frac{4r_0}{L_{\mathrm j}}\Delta
X_{0}^{2}(x-1)}}+\frac{\Phi \,\Delta X_{0}}{1-\frac{ \Gamma\,\Delta X_{0}}{\sqrt{8\pi}}\frac{r_0}{D} (x-1)}.
\end{equation}
The corrections due to nonlinear wave steepening are small as long as $\Delta X\gtrsim 10^{-2}$.  In those cases, we only need to consider the first term and hence retain the earlier solution which we derive in Sect.~\ref{sect:a} and display in Figs.~\ref{figure3} and \ref{figure4}.

\section{Conclusions }

We have shown that traveling solar-wind bulk-velocity jumps effectively process solar-wind electrons in energy at their propagation outwards from the Sun through the inner heliosphere. These fluctuations in the solar-wind bulk velocity are ubiquitous as shown in Fig.~\ref{figure1}. In an earlier paper, we have shown that this energization can be expressed in terms of a substantial temperature increase of the solar-wind
electrons at larger distances from the Sun of about 50 AU to 90 AU. Assuming that
the jump amplitude $\Delta X=\Delta U/U$ of the propagating bulk-velocity
structure does not change with solar distance $r$, the previous study predicts electron
temperatures of more than  $10^{5}$ K at 90 AU (i.e., at about position of the solar-wind termination shock). In this study, we introduce higher-order corrections due to the fact that the energy for the energization of solar-wind electrons is taken from the kinetic excess energy of the
propagating jump structure. We find that the previous assumption of a constant jump amplitude $\Delta X$ is most probably unrealistic. In addition, such jump structures do
permanently work against the ion excess pressure on the downstream side of
the shock structure. Taking into account these two physical processes allows us to quantitatively show how the jump amplitude $\Delta X=\Delta X(r)$ decreases with distance from the Sun, eventually reducing $\Delta X$ independent of the initial value $\Delta X_{0}$ of the jump amplitude to
values of $\lesssim 0.1$ at the termination shock as shown in Fig.~\ref{figure2}. The nonlinear pile-up of bulk-velocity fluctuations counteracts these two mechanisms with the tendency to reform the solitary jump structure by forming waves at larger scales. We find, however, that this mechanism is most likely not effective enough to compensate for the reduction of $\Delta X$ with distance, unless $\Delta X\lesssim 10^{-2}$.

Although the described mechanisms lead to a reduction of $\Delta X$ with distance from the Sun as shown in Fig.~\ref{figure2}, the jump-induced heating mechanism still leads to higher electron temperatures than anticipated due to adiabatic cooling at solar distances beyond 10 AU. We predict values above 6000 K to 20\,000 K (strongly depending on the initial value of the jump amplitude $\Delta X_{0}$ in the innermost heliosphere at $r=r_{0}=1\,\mathrm{AU}$) at distances beyond 50 AU with the solar-wind electron-temperature profiles $T_{\mathrm e}(r)$ shown in Figs.~\ref{figure3} and \ref{figure4}. Our results show a very good agreement with in-situ measurements of the electron-core temperature in the plane of the ecliptic from the Ulysses spacecraft. We achieve the best agreement for values of $\Delta X_0$ between 0.25 and 0.3, suggesting that these values describe the realistic initial jump amplitude in the plane of the ecliptic. In-situ observations of the electron temperature are not available for heliocentric distances beyond 5 AU, so that our results are a predictive extrapolation beyond the explored range. 

We conclude that solar-wind electrons do not rapidly cool off with distance from the Sun as it has been generally assumed up to now. They cannot be considered cold beyond 10 AU. Instead, they need to be considered keeping track with the solar wind ion temperatures at large distances \citep[see][]{richardson95}.

\begin{acknowledgements}
D.V.~is supported by NASA grant NNX12AB27G. We used data from NASA's OMNIWeb Service provided by the Goddard Space Flight Center Space Physics Data Facility (GSFC/SPDF), as well as Ulysses data provided by NASA's National Space Science Data Center (NSSDC).  
\end{acknowledgements}

\bibliographystyle{aa}
\bibliography{electron_heating}

\end{document}